\documentclass[letterpaper,10pt,twocolumn,oneside]{article}

\RequirePackage{./endfloat/endfloat}

\usepackage{geometry}
\usepackage[english]{babel}

 \usepackage{amsmath}
 \usepackage{amsfonts}

\usepackage[T1]{fontenc}
\usepackage{lmodern}

\usepackage{amssymb}
\usepackage{graphicx}
\usepackage{dsfont} 

\usepackage{color}
\usepackage{array}
\usepackage{supertabular}
\usepackage{hhline}
\usepackage{multirow}
\usepackage[table]{xcolor}
\usepackage{amsthm}

\usepackage{float}

\usepackage{longtable}

\usepackage{stfloats}

\usepackage[noprefix]{nomencl} 

\makenomenclature

\newcommand{\refeq}[1]{(\ref{#1})}

\makeatletter

\@addtoreset{equation}{section}
\makeatother

\theoremstyle{plain}
\newtheorem{Th}{Theorem}[section]

\newtheorem{Lem}{Lemma}[section]

\newtheorem{Def}{Definition}[section]

\theoremstyle{definition}

\newtheorem{Rmq}{Remark}[section]

\let\oldqedsymbol=\qedsymbol

\renewcommand{\qedsymbol}{\oldqedsymbol}

\newenvironment{pfof}[1]{\vspace{1ex}\noindent{\bf Proof of #1:}\hspace{0.5em}}
	{\hfill\qed\vspace{1ex}}

\begin{document}

\twocolumn[
\begin{center}
{\Large
	{\sc A moving fixed-interval filter/smoother for estimation of vehicle position using odometer and map-matched GPS}
}
\bigskip

 Cindie Andrieu $^{1}$, Guillaume Saint Pierre $^{1}$ and Xavier Bressaud $^{2}$
\bigskip

{\it
$^{1}$ IFSTTAR, IM, LIVIC, 14, route de la Minière, 78000 Versailles, France.\\
E-mail : cindie.andrieu@ifsttar.fr and guillaume.saintpierre@ifsttar.fr

$^{2}$ Université Paul Sabatier, Institut de Mathématiques de Toulouse, F-31062 Toulouse Cedex 9, France. E-mail: bressaud@math.univ-toulouse.fr
}
\end{center}
\bigskip
]


{\bf Abstract.}
\smallskip
This paper presents some optimal real-time and post-processing estimators of vehicle position using odometer and map-matched GPS measurements. These estimators were based on a simple statistical error model of the odometer and the GPS which makes the model generalizable to other applications. Firstly, an asymptotically minimum variance unbiased estimator and two optimal moving fixed interval filters which are more flexibles are exposed. Then, the post-processing case leads to the construction of two moving fixed interval smoothers. These estimators are tested and compared with the classical Kalman filter with simulated and real data, and the results show a good accuracy of each of them.


\section{Introduction}

The development of Field Operational Test (FOT) and Naturalistic Driving Study (NDS) allow to collect large databases that provide a wealth of information regarding driving behavior and more generally the interactions between driver, vehicle and/or environment factors (e.g. the SHRP 2 NDS with about 3000 vehicles in the United States for 2 years \cite{Shrp2_2010}, and the EuroFOT project with about 1000 vehicles in Europe for 1 year \cite{Eurofot2009}). These mass data, generally collected from Floating Car Data (FCD), can be used both to study global effects by calculating aggregated indicators such as mean or median, and also effects at a more local scale by studying individual speed or accelerate profiles. Some studies have shown that space-speed profiles (speed versus vehicle position) are very informative to study driver behavior and the effects of some infrastructure elements (for example, behavioral studies at a signalised intersection \cite{Laureshyn2005}, or effects of traffic calming measures such as speed humps and speed cushions \cite{Barbosa2000}). Such studies require relatively accurate location information. \\
Global navigation satellite systems (GNSS), such as the Global Positioning System (GPS), are commonly used for vehicle positioning and are based on measurements of the propagation time of a signal between each visible satellites and the receiver. However, GNSS performance is highly dependent on the environment, and in urban environments the signal is affected by many errors due to satellite masking and multipath. A common solution is to use additional sensors such to overcome the weaknesses of GNSS. In practice, the reliability of vehicle positioning is obtained by the coupling of GNSS that provide absolute positioning, with dead reckoning (DR) system, such as odometer and gyroscope, that provides vehicle's position relative to an initial position \cite{Lahrech2004,Kealy1999,Zhao2003}. However, in the long term the performance of DR systems is poor due to the accumulation of measurements errors over time. Thus, positioning information from GNSS and DR systems are complementary. \\
Many methods exist for multi-sensor vehicle navigation (e.g. neural networks \cite{Chiang2001}, fuzzy logic \cite{Loebis2004}, particle filter \cite{Boucher2004}) but Kalman filtering/smoothing techniques are the most used for their speed and ease of implementation \cite{Kealy1999,Ramjattan1995,Da1995,Lahrech2004,Zhao2003}. The Kalman filter/smoother is a recursive algorithm to estimate a signal from noisy measurements, based on a compromise between a predictive dynamic model and a measurement model. The Kalman filter is a real-time estimator that uses only the past observations $y(k)\ (0\leq k \leq t)$ to estimate the state vector $x(t)$ at the time $t$. The basic Kalman filter (\cite{Kalman1960}), based on least squares approach, is an optimal estimator under the assumptions of linearity of the system and the gaussian distribution of the errors. Some extensions algorithms have also been developed, such as the Extended Kalman Filter (EKF) and the Unscented Kalman Filter (UKF), in the case of nonlinear systems. However, in Naturalistic Driving Studies, data are usually post-processed and it is desirable to dispose all the measurement data of the experiment in order to achieve better estimation accuracy. Estimators that take into account both past and future observations are often called smoothers. Fixed-Interval Smoothing (FIS) algorithms, based on Kalman filtering/smoothing theory, involve measurements over a given fixed time interval $[0,T]$ and use all the measurements $y(k)\ (0\leq k \leq T,\ T>t)$ to estimate the state vector $x(t)$. Fixed-interval smoothers are generally two-filter smoothers based on a combination of a forward and a backward estimate : a forward pass that processes a Kalman filter, and a backward pass that operates backward in time by using the measurements after the time t. The most popular fixed-interval smoothing algorithms are the Rauch-Tung-Striebel (RTS) smoother \cite{Rauch1965}, the Main-Fraser smoother \cite{Mayne1966,Fraser1967} and the Wall-Willsky-Sandell smoother \cite{Wall981}. The main drawback of these smoothers is that they require the operation of two filters.\\
This paper presents some optimal real-time and post-processing estimators of the distance traveled by a vehicle on a road segment relative to an initial position, using odometer and map-matched GPS measurements. The main contributions of this paper are to propose a simple error model of the sensors which makes the model generalizable to other applications while being efficient, and to propose moving fixed interval filter/smoother which allow flexibility of use. In section 2, the statistical model is explained and the construction of the estimators are developed. Firstly, two real-time estimators are exposed: an asymptotically minimum variance unbiased estimator and an optimal moving fixed interval filter. Then, a generalization of the two previous filters in the post-processing case, leads to the construction of two moving fixed interval smoothers. The effectiveness of these estimators is tested and a comparison with the Kalman filter is performed in section 3 with simulated and real data. Finally, a discussion about the results is proposed.

\section{Methodology}

\subsection{Statistical modelisation}
\label{subsection_statistical_modelisation}

The aim of this study is to estimate the vehicle position $x(t_{i})$ at time $t_{i}$ on a road segment. We denote $\{X(t): t\in[0,T]\}$ the continuous random process representing the vehicle position on the time interval $[0,T]$, and $\{X(t_{i}):i=1,...,n\}$ the sampled process. Let $\{x(t_{1}),...,x(t_{n})\}$ a realization of this random process. Our aim is to estimate this realization from odometer and GPS noisy data.\\
Let $n$ and $m$, two integers with $m\leq n$, the number of measurements respectively provided by the odometer and the GPS. In the remainder of this paper, it is assumed that the GPS measurements are map-matched, so that the vehicle is positioned on the correct road segment. Many map-matching algorithms have been developed to identify the correct road segment on which the vehicle is travelling. Most of these algorithms use navigation data from GPS and digital spatial road network data and current map-matching algorithms are described in \cite{Quddus2007}, but the choice of the correct road segment is not the subject of this study. We suppose that the correct road segment have been identified and we search to determine the vehicle location on that segment. For example, Taylor et al. (2006) developed in \cite{Taylor2006} a map-matching algorithm called OMMGPS that combine GPS pseudorange observations and odometer positions to provide a vehicle position at 1s epochs. \\
In this study, the map-matched GPS measurements denoted $(y_{gps}(t_{0}),\ldots,y_{gps}(t_{m}))$ represent the curvilinear abscissa of the vehicle on the studied road segment (absolute location). GPS position data are affected by many errors including atmospheric and ionospheric errors, satellite orbit errors, satellite clock errors, and multipath errors. We represent these errors by a white Gaussian noise which is a classical hypothesis especially in Kalman filtering (\cite{Grewal2007}, \cite{Scott1995}), even if current studies have shown that noises are non centered Gaussian distributions in urban environments but rather Gaussian mixture (\cite{Viandier2010}). The odometer measurements denoted $(y_{od}(t_{0}),\ldots,y_{od}(t_{n}))$ represent the distance traveled by the vehicle from the initial position $x(t_{0})$ (relative location). Odometer data are affected by many errors which are divided into two categories: systematic errors related to the properties of the vehicle (mainly unequal wheel diameters and uncertainty about the wheelbase) and nonsystematic errors related to the environment (mainly wheel slippage due to slippery roads, over-acceleration, ...)(\cite{Borenstein1996}). Nonsystematic errors are very difficult to estimate because any unexpected irregularity can introduce a huge error, while systematic errors accumulate constantly over time. In our study, we propose a simple modeling of odometer errors and we represented them by a cumulative sum of centered Gaussian distributions. So the discretized observation model of these two sensors can be written as the following system:
\begin{equation}
\label{eq_model_observation}
\left\{
\begin{array}{l}
y_{od}(t_{i})=x(t_{i})-x(t_{0})+\sum_{k=1}^{i} \varepsilon_{od,k}\ \\\text{\hspace{3em}with}\ t_{i}=\frac{iT}{n},\ i=1,...,n \\
y_{gps}(t'_{j})=x(t'_{j})+ \varepsilon_{gps,j}\ \\\text{\hspace{3em}with}\ t'_{j}=\frac{jT}{m},\ j=1,...,m
\end{array}
\right.
\end{equation}
where $\varepsilon_{od,i}$ and $\varepsilon_{gps,j}$ are independent gaussian centered errors with respective variance $\sigma_{od}^{2}$ and $\sigma_{gps}^{2}$. \\
To simplify the model, we assume that the initial position $x(t_{0})$ is zero. Thus, the odometer model and the GPS model described in \refeq{eq_model_observation} differ only by measurement errors and sampling rate: a high sampling rate with accumulating errors for the odometer, and generally a lower sampling rate without accumulating errors for the GPS. The construction of an estimator $\widehat{x}(t_{i})$ of the vehicle position $x(t_{i})$ at the sampling time $t_{i},\ i=1,\ldots,n$ from noisy measurements of GPS and odometer will take into account advantages and disadvantages of these two sensors. \\
Later in the paper, we will denote $\lambda=\frac{f_{od}}{f_{gps}}$ the ratio between the odometer and GPS sampling frequencies (in practice, $\lambda \geq 1$) and we will assume that $\lambda \in \mathds{N}^{*}$, i.e. that for some time $t_{i}$ we have both odometer and GPS measurements.

\subsection{Real-time estimator}
\label{subsection_real_time_estimator}

In this section, vehicle position is estimated in real-time, i.e. the position $x(t_{i})$ at a given sampling time $t_{i}$ is estimated using only measurements obtained up to time $t_{i}$.

\subsubsection{Asymptotically minimum variance unbiased estimator}

The main idea is to use odometer measurements, which has the advantage of having a high sampling rate and provide good accuracy in the short term, and to readjust with the GPS measurements, when they are available, to compensate for the accumulation of positional errors. Our estimator is then defined as follows:
\begin{Def}
\label{def_estim_RT_asymp_recurs}
Let $\lambda=\frac{f_{od}}{f_{gps}} \in \mathds{N}^{*}$. The estimator $\widehat{x}^{\infty}_{RT}(t_{i})$ is a real-time estimator of the vehicle position at the given sampling time $t_{i},\ i=1,\ldots,n$, defined recursively as follows: For $i=1,\ldots,n$,


\begin{equation}
\label{eq_estim_RT_asymp_recurs}
\left\{
\begin{array}{l}
\widehat{x}^{\infty}_{RT}(t_{i})=w_{1}\ [\widehat{x}^{\infty}_{RT}(t_{i-1})+y_{od}(t_{i})-y_{od}(t_{i-1})]\\
           \text{\hspace{1em}}                     +w_{2}\ y_{gps}(t_{i})\ \ \text{if}\ \ i\equiv 0\ (mod\ \lambda) \\
\text{Otherwise,}\\
\widehat{x}^{\infty}_{RT}(t_{i})=\widehat{x}^{\infty}_{RT}(t_{i-1})+y_{od}(t_{i})-y_{od}(t_{i-1})\ \
\end{array}
\right.
\end{equation}

where $w_{1}+w_{2}=1$.
\end{Def}

\noindent
In practice, the initial position is unknown, so we suppose that: \\
$\widehat{x}^{\infty}_{RT}(t_{0})=
\left\{
\begin{array}{l}
y_{gps}(t_{0})\ \ \text{if}\ \ y_{gps}(t_{0})\ \ \text{is available} \\
y_{od}(t_{0})\ \ \text{otherwise}
\end{array}
\right.$ \\

\begin{Th}
\label{th_estim_RT_asymp_recurs_weight_and_var}
The real-time estimator $\widehat{x}^{\infty}_{RT}$ defined in definition \ref{def_estim_RT_asymp_recurs} with the following weights: \\
$w_{1}=\frac{\lambda\,r+2-\sqrt{\lambda\,r(\lambda\,r+4)}}{2}\ \text{and}\ w_{2}=\frac{-\lambda\,r+\sqrt{\lambda\,r(\lambda\,r+4)}}{2}$
where $r=\frac{\sigma^{2}_{od}}{\sigma^{2}_{gps}}$ is the ratio between the odometer and GPS variances, is an asymptotically minimum variance unbiased estimator. The asymptotic variance can be written as follows:
\begin{equation}
\label{eq_var_estim_RT_asymp_recurs}
Var[\widehat{x}^{\infty}_{RT}(t_{i})] \rightarrow \sigma^{2}_{od}\ \frac{\lambda\ w^{2}_{1}+\frac{1}{r}w^{2}_{2}}{1-w^{2}_{1}} \ \ \text{when}\ \ t_{i} \rightarrow \infty
\end{equation}
\end{Th}

The recursive definition of the estimator given in definition \ref{def_estim_RT_asymp_recurs} has the advantage of being simple to compute. However, in general, recursive algorithms require more computational resource than iterative algorithms. So, we give a non-recursive expression of the real-time estimator with asymptotically minimum variance $\widehat{x}^{\infty}_{RT}$ defined in definition \ref{def_estim_RT_asymp_recurs}.

\begin{Def}
\label{def_estim_RT_asymp_nonrecurs}
Let $\lambda=\frac{f_{od}}{f_{gps}} \in \mathds{N}^{*}$ and $\lfloor x \rfloor$ the floor function. The estimator $\widehat{x}^{\infty}_{RT}(t_{i})$ is a real-time estimator, with asymptotically minimum variance, of the vehicle position at the given sampling time $t_{i},\ i=1,\ldots,n$, defined as follows: \\
\begin{equation}
\label{eq_estim_RT_asymp_nonrecurs}
\widehat{x}^{\infty}_{RT}(t_{i})=\sum_{j=1}^{N}\widetilde{w}_{j}^{-}\widehat{x}_{j}^{-}(t_{i})\ \ \ \text{with}\ \ N=\lfloor\frac{i}{\lambda}\rfloor+1
\end{equation}
where for $j=1,...,N$,
\begin{equation}
\label{eq_estim_moins}
\begin{split}
\widehat{x}^{-}_{j}(t_{i})&=y_{gps}(t_{g^{-}_{i}(j)})\\
&\quad +\sum_{k=g^{-}_{i}(j)+1}^{i}(y_{od}(t_{k})-y_{od}(t_{k-1})) \ \ \\
\text{with}&\  \ g^{-}_{i}(j)=\lambda\lfloor\frac{i}{\lambda}\rfloor-\lambda(j-1)
\end{split}
\end{equation}

and the weights $\widetilde{w}_{j}^{-}$ are defined by
$\left\{
\begin{array}{l}
\widetilde{w}_{j}^{-}=w_{2}w_{1}^{j-1}\ \ \text{if}\ j<N \\
\widetilde{w}_{N}^{-}=w_{1}^{N-1}
\end{array}
\right.$\\
with $(w_{1},w_{2})$ the weights defined in theorem \ref{th_estim_RT_asymp_recurs_weight_and_var}.
\end{Def}

The equivalence with the recursive expression of the estimator $\widehat{x}^{\infty}_{RT}$ given in definition \ref{def_estim_RT_asymp_recurs} is easily demonstrated by recursion.

\begin{Rmq}
~\\
\begin{enumerate}
  \item It is easy to check that the sum of weights $\widetilde{w}_{j}^{-}$ is equal to one.
  \item For a given sampling time $t_{i}$, the estimators $\widehat{x}^{-}_{j}(t_{i}),\ j=1,\ldots,N$, are also estimators of the vehicle position at time $t_{i}$, each estimator being associated with the j-th GPS measurement obtained before time $t_{i}$ as shown in Figure \ref{fig_estim_moins_ENG}. The real-time estimator $\widehat{x}^{\infty}_{RT}$ is a weighted sum of these estimators.

\begin{figure}[H]
\begin{center}
\includegraphics[width=10cm]{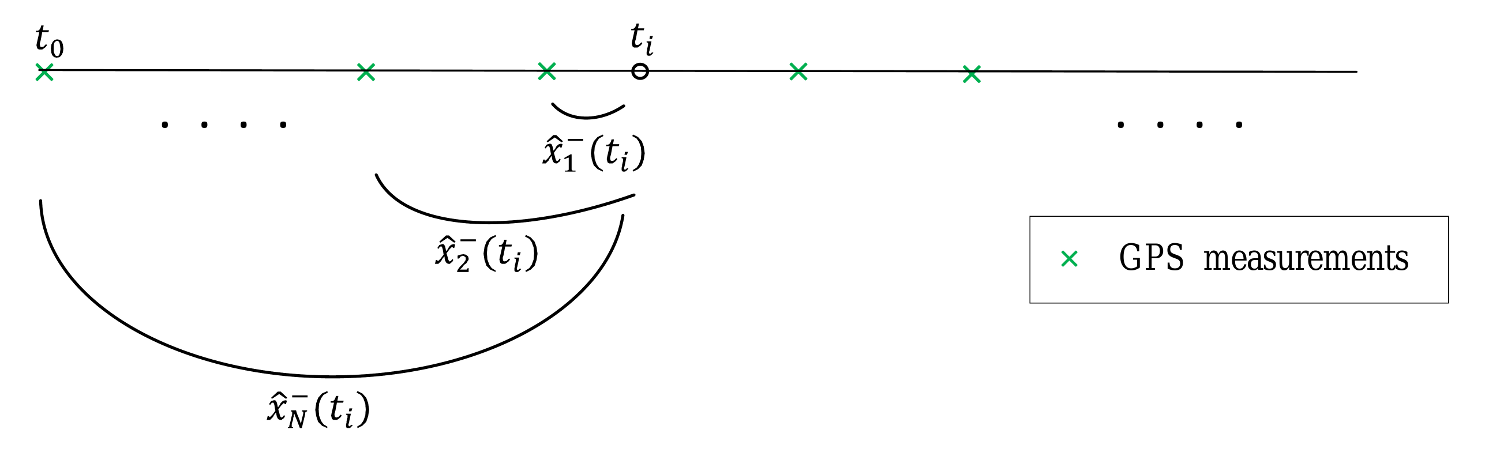}
\end{center}
\caption{Graph of estimators $\widehat{x}^{-}_{j},\ j=1,\ldots,N$.}
\label{fig_estim_moins_ENG}
\end{figure}

\end{enumerate}
\end{Rmq}

The estimator $\widehat{x}^{\infty}_{RT}$ defined in both \refeq{def_estim_RT_asymp_recurs} and \refeq{eq_estim_RT_asymp_nonrecurs} uses all measurements obtained up to time $t_{i}$. However, with the non-recursive expression \refeq{eq_estim_RT_asymp_nonrecurs}, it is possible to fix an integer $N < \lfloor\frac{i}{\lambda}\rfloor+1$ in order to obtain a "truncated" estimator that can be more advantageous to calculate from a computational point of view. In this case, the integer N represents the number of GPS measurements (available before the time $t_{i}$) used in the computation of the estimator $\widehat{x}^{\infty}_{RT}$. The choice of the value of N is entirely defined by the user which implies a high flexibility in practice. We then deduce an expression of the variance of the estimator $\widehat{x}^{\infty}_{RT}$ with N fixed, at each sampling time $t_{i}$.

\begin{Th}
\label{th_var_estim_RT_asymp_nonrecurs}
Let $N \geq 1$ an integer, $r=\frac{\sigma^{2}_{od}}{\sigma^{2}_{gps}}$ and $\lambda=\frac{f_{od}}{f_{gps}} \in \mathds{N}^{*}$. Let $\widehat{x}^{-}_{j},\ j=1,\ldots,N$ and $\widetilde{w}_{j}^{-},\ j=1,\ldots,N$ respectively the estimators and the weights defined in definition \ref{def_estim_RT_asymp_nonrecurs}. Then the variance of the real-time estimator $\widehat{x}^{\infty}_{RT}$ at a sampling time $t_{i},\ i=1,\ldots,n$ is written in matrix form as follows:
\begin{equation}
\label{eq_var_matrix_estim_RT_asymp_nonrecurs}
Var[\widehat{x}^{\infty}_{RT}(t_{i})] = (\mathbf{\widetilde{w}^{-}})^{T}\, \mathbf{\Sigma^{-}} \mathbf{\widetilde{w}^{-}}
\end{equation}
where $\mathbf{\widetilde{w}^{-}}=(\widetilde{w}_{1}^{-},\ldots,\widetilde{w}_{N}^{-})^{T}$ and $\mathbf{\Sigma^{-}}$ is the $N\times N$ covariance matrix of the estimators $\widehat{x}^{-}_{j}$. The covariance matrix $\mathbf{\Sigma^{-}}$ can be decomposed as follows:
\begin{equation}
\label{eq_covariance_matrix_sigma_moins}
\mathbf{\Sigma^{-}}=\sigma^{2}_{gps}(\mathbf{I_{N}}+r \mathbf{A_{N}}(d_{i}))
\end{equation}
where $\mathbf{I_{N}}$ is the identity matrix of size $N$, $d_{i}=i-\lambda\lfloor\frac{i}{\lambda}\rfloor$ is the number of odometer measurements between $t_{i}$ and the first time of a GPS measurement before $t_{i}$, and $\mathbf{A_{N}}(d_{i})$ is a $N\times N$ matrix, function of $d_{i}$, defined by:

\small

\begin{equation}
\label{eq_matrix_A}
\begin{split}
\mathbf{A_{N}}(d_{i})=\begin{bmatrix}
d_{i} & d_{i} & d_{i} & \cdots & d_{i} \\
d_{i} & d_{i}+\lambda & d_{i}+\lambda & \cdots & d_{i}+\lambda \\
d_{i} & d_{i}+\lambda & d_{i}+2\lambda & \cdots & d_{i}+2\lambda \\
\vdots & \vdots & \vdots & \ddots & \vdots \\
d_{i} & d_{i}+\lambda & d_{i}+2\lambda & \cdots & d_{i}+(N-1)\lambda
\end{bmatrix}\\
\vspace{1ex}
\end{split}
\end{equation}

\vspace{3ex}

\normalsize

Then, we can also deduce a linear form of the variance of the real-time estimator $\widehat{x}^{\infty}_{RT}$ at a sampling time $t_{i},\ i=1,\ldots,n$ as follows:

\begin{equation}
\begin{split}
\label{eq_var_linear_estim_RT_asymp_nonrecurs}
Var[\widehat{x}^{\infty}_{RT}(t_{i})]&=\sigma_{gps}^{2}[\frac{(1-w_{1})^{2}+2w_{1}^{2N-1}(1-w_{1})}{1-w_{1}^{2}} \\
&+ r(d_{i}+\lambda \frac{w_{1}^{2}-w_{1}^{2N}}{1-w_{1}^{2}})]
\end{split}
\end{equation}

where $w_{1}=\frac{\lambda\,r+2-\sqrt{\lambda\,r(\lambda\,r+4)}}{2}$ is the asymptotically optimal weight defined in theorem \ref{th_estim_RT_asymp_recurs_weight_and_var}.
\end{Th}

Note that for a fixed N, the variance function defined in \refeq{eq_var_linear_estim_RT_asymp_nonrecurs} is periodic with period $\lambda$. Moreover, since $\widehat{x}^{\infty}_{RT}$ is an asymptotically minimum variance estimator, the weights $\widetilde{w}_{j}^{-},\ j=1,\ldots,N$, defined in definition \ref{def_estim_RT_asymp_nonrecurs} are optimal when $N$ tends to infinity, i.e. when we have an infinite number of measurements. Thus, if we assume $d_{i}=0$ and if $N$ tends to infinity in the variance expression \refeq{eq_var_linear_estim_RT_asymp_nonrecurs}, we find the expression of the asymptotically variance given in \refeq{eq_var_estim_RT_asymp_recurs}. The speed of convergence of the variance of $\widehat{x}^{\infty}_{RT}$ defined in \refeq{eq_var_linear_estim_RT_asymp_nonrecurs} to the asymptotically variance defined in \refeq{eq_var_estim_RT_asymp_recurs} will be examined in the section \ref{section_data_processing_and_discussion}.\\
However, the estimator $\widehat{x}^{\infty}_{RT}$ is an asymptotically minimum variance estimator and it is not optimal for estimating the vehicle position at a sampling time $t_{i}$ close to the initial time $t_{0}$. Therefore we have also constructed an optimal real-time estimator with minimum variance at each sampling time $t_{i}$.


\subsubsection{Minimum variance unbiased estimator for a fixed N}

In the definition \ref{def_estim_RT_asymp_nonrecurs}, we have written a real-time estimator of the vehicle position as a weighted average of the estimators $\widehat{x}^{-}_{j}$ ($j=1,\ldots,N$) and we have determined the optimal weights $\widetilde{w}_{j}^{-}$ ($j=1,\ldots,N$) that minimize the asymptotic variance. In this section, we consider the same real-time estimator but we search the optimal weights $w_{j}^{-}$ ($j=1,\ldots,N$) that minimize the variance at each sampling time $t_{i}$.

\begin{Th}
\label{th_estim_RT_opt}
Let $N \geq 1$ an integer and $\mathbf{b}=(1,\ldots,1)^{T}$ a vector of length N. Assume that $\widehat{x}^{-}_{j},\ j=1,\ldots,N$ are the estimators defined in definition \ref{def_estim_RT_asymp_nonrecurs} and $\mathbf{\Sigma^{-}}$ is the $N\times N$ covariance matrix of these estimators defined in theorem \ref{th_var_estim_RT_asymp_nonrecurs}. The estimator $\widehat{x}^{opt}_{RT}(t_{i})$ is a real-time estimator of the vehicle position at the given sampling time $t_{i},\ i=1,\ldots,n$, defined as follows: \\
\begin{equation}
\label{eq_estim_RT_opt}
\widehat{x}^{opt}_{RT}(t_{i})=\sum_{j=1}^{N} \widehat{w}_{j}^{-}\widehat{x}_{j}^{-}(t_{i})
\end{equation}
where the weight vector $\mathbf{\widehat{w}^{-}}=(\widehat{w}^{-}_{1},\ldots,\widehat{w}^{-}_{N})^{T}$ satisfies:
\begin{equation}
\label{eq_poids_estim_RT_opt}
\begin{split}
\mathbf{\widehat{w}^{-}}&=\frac{1}{c_{rt}}(\mathbf{\Sigma^{-}})^{-1}\mathbf{b}\ \ \\
\text{with}&\ \ c_{rt}=\mathbf{b}^{T}(\mathbf{\Sigma^{-}})^{-1}\mathbf{b}\ \text{a constant.}
\end{split}
\end{equation}
Then, $\widehat{x}^{opt}_{RT}(t_{i})$ is a minimum variance unbiased estimator of the vehicle position at the sampling time $t_{i}$, and its variance at time $t_{i}$ is the following:
\begin{equation}
\label{eq_var_estim_RT_opt}
Var[\widehat{x}_{RT}^{opt}(t_{i})]=\frac{1}{c_{rt}}
\end{equation}
\end{Th}

The variance function defined in \refeq{eq_var_estim_RT_opt} depends on $d_{i}$ and is periodic with period $\lambda$. The estimator $\widehat{x}^{opt}_{RT}$ is a minimum variance unbiased estimator at each sampling time $t_{i}$ for a fixed N. However, determining the optimal weights $\widehat{w}_{j}^{-}$ requires the inversion of the covariance matrix $\mathbf{\Sigma^{-}}$ which is inconvenient in practice. Furthermore, the expressions of the optimal weights and the variance of $\widehat{x}^{opt}_{RT}$ are not given explicitly in terms of the integer N, which makes it difficult to study the properties of this estimator depending on N. Thus, in some cases, it may be more advantageous to use the "truncated" estimator $\widehat{x}^{\infty}_{RT}$ with a fixed N that have a simpler expression of weights and variance.


\subsection{Post-processing estimator}
\label{subsection_post_processing_estimator}

In this section, we assume that data are post-processed and we can use all the GPS and odometer measurements available on the studied time interval $[0,T]$. Thus, unlike the previous section where we were restricted to use only measurements obtained up to time $t_{i}$ to estimate the vehicle position at the sampling time $t_{i}$, the objective of this section is to use all available information to construct a more accurate estimator than the real-time estimators defined in the section \ref{subsection_real_time_estimator}. \\

\subsubsection{Minimum variance unbiased estimator for a fixed N}

The general idea is to extend the real-time estimator defined in definition \ref{def_estim_RT_asymp_nonrecurs} in case we also have measurements obtained after the sampling time $t_{i}$.

\begin{figure}[H]
\begin{center}
\includegraphics[width=10cm]{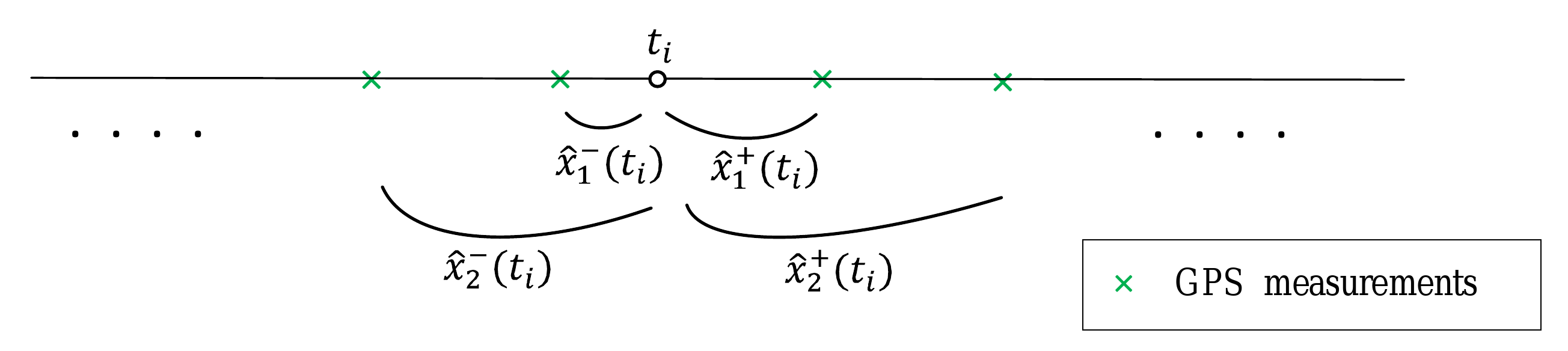}
\end{center}
\caption{Graph of estimators $\widehat{x}^{-}_{j}$ and $\widehat{x}^{+}_{j},\ j=1,\ldots,N$.}
\label{fig_estim_moins_et_plus_ENG}
\end{figure}

\begin{Def}
\label{def_estim_PP}
Let $\lambda=\frac{f_{od}}{f_{gps}} \in \mathds{N}^{*}$ and $\lfloor x \rfloor$ the floor function. Assume that $N\geq 1$ is a fixed integer. The estimator $\widehat{x}_{PP}(t_{i})$ is a post-processing estimator of the vehicle position at the given sampling time $t_{i},\ i=1,\ldots,n$, defined as follows: \\

\begin{equation}
\label{eq_estim_PP}
\begin{split}
\widehat{x}_{PP}(t_{i})&=\sum_{j=1}^{N}(w^{-}_{j}\widehat{x}^{-}_{j}(t_{i})+w^{+}_{j}\widehat{x}^{+}_{j}(t_{i}))
\ \ \\
\text{with}&\ \ \sum_{j=1}^{N}(w^{-}_{j}+w^{+}_{j})=1
\end{split}
\end{equation}

where for $j=1,...,N$,

\begin{equation}
\label{eq_estim_moins}
\begin{split}
\widehat{x}^{-}_{j}(t_{i})&=y_{gps}(t_{g^{-}_{i}(j)})\\
& \quad +\sum_{k=g^{-}_{i}(j)+1}^{i}(y_{od}(t_{k})-y_{od}(t_{k-1})) \ \ \\
\text{with}&\ \ g^{-}_{i}(j)=\lambda\lfloor\frac{i}{\lambda}\rfloor-\lambda(j-1)
\end{split}
\end{equation}

and

\begin{equation}
\label{eq_estim_plus}
\begin{split}
\widehat{x}^{+}_{j}(t_{i})&=y_{gps}(t_{g^{+}_{i}(j)})\\
&\quad -\sum_{k=i+1}^{g^{+}_{i}(j)}(y_{od}(t_{k})-y_{od}(t_{k-1}))
\ \ \\
\text{with}&\ \ g^{+}_{i}(j)=\lambda\lfloor\frac{i}{\lambda}\rfloor+\lambda\,j
\end{split}
\end{equation}

A graph of the estimators $\widehat{x}^{-}_{j}$ and $\widehat{x}^{+}_{j}$ is represented in Figure \ref{fig_estim_moins_et_plus_ENG}.
\end{Def}

In this case, the integer N represents the number of GPS measurements (available before and after the time $t_{i}$) used in the computation of the estimator (i.e. a total of $2N$ GPS measurements around $t_{i}$).

The following lemma gives a general expression of the variance of the post-processing estimator defined in \refeq{eq_estim_PP}.

\begin{Lem}
\label{lemme_var_estim_PP}
Let $N \geq 1$ an integer and $\widehat{x}^{-}_{j}$ and $\widehat{x}^{+}_{j}$, $j=1,\ldots,N$, the estimators defined in definition \ref{def_estim_PP}. Let $\mathbf{\widehat{w}}=(\widehat{w}^{-}_{1},\ldots,\widehat{w}^{-}_{N},\widehat{w}^{+}_{1},\ldots,\widehat{w}^{+}_{N})^{T}$ the weight vector of length $2N$. The variance of the post-processing estimator $\widehat{x}_{PP}$ defined in definition \ref{def_estim_PP} can be written as follows:
\begin{equation}
\label{eq_var_estim_PP}
Var[\widehat{x}_{PP}(t_{i})]=\mathbf{w}^{T}\mathbf{\Sigma}\,\mathbf{w}
\end{equation}
where $\mathbf{\Sigma}$ is the $2N\times 2N$ covariance matrix of the estimators $\widehat{x}^{-}_{j}$ and $\widehat{x}^{+}_{j}$ defined by
$\Sigma=
\begin{bmatrix}
\Sigma^{-}&0 \\
0&\Sigma^{+}
\end{bmatrix}$
with $\mathbf{\Sigma^{-}}$ and $\mathbf{\Sigma^{+}}$ respectively the $N\times N$ covariance matrix of the estimators $\widehat{x}^{-}_{j}$ and $\widehat{x}^{+}_{j}$. Furthermore, we can decomposed $\mathbf{\Sigma^{-}}$ and $\mathbf{\Sigma^{+}}$ as follows:

\begin{equation}
\label{eq_covariance_matrix_sigma_moins}
\begin{split}
&\mathbf{\Sigma^{-}}=\sigma^{2}_{gps}(\mathbf{I_{N}}+r \mathbf{A_{N}}(d_{i}))\ \ \text{and}\ \ \\ &\mathbf{\Sigma^{+}}=\sigma^{2}_{gps}(\mathbf{I_{N}}+r \mathbf{A_{N}}(\lambda-d_{i}))
\end{split}
\end{equation}

where $\mathbf{I_{N}}$ is the identity matrix of size $N$, $d_{i}=i-\lambda\lfloor\frac{i}{\lambda}\rfloor$ is the number of odometer measurements between $t_{i}$ and the first time of a GPS measurement before $t_{i}$, and $\mathbf{A_{N}}(d_{i})$ is a $N\times N$ matrix, function of $d_{i}$, defined in \refeq{eq_matrix_A}.
\end{Lem}

Then, we search the optimal weights $(w_{j}^{-},w_{j}^{+})$ ($j=1,\ldots,N$) that minimize the variance of the post-processing estimator $\widehat{x}_{PP}$ at each sampling time $t_{i}$. Intuitively, we give more weight to the estimators $\widehat{x}^{-}_{j}(t_{i})$ and $\widehat{x}^{+}_{j}(t_{i})$ associated with GPS measurements obtained at times close to $t_{i}$. The minimum variance unbiased estimator $\widehat{x}^{opt}_{PP}(t_{i})$ of the vehicle position at the sampling time $t_{i}$ is similar to the minimum variance unbiased estimator $\widehat{x}^{opt}_{RT}(t_{i})$ defined in theorem \ref{th_estim_RT_opt} by taking $\mathbf{\widehat{w}}=(\widehat{w}^{-}_{1},\ldots,\widehat{w}^{-}_{N},\widehat{w}^{+}_{1},\ldots,\widehat{w}^{+}_{N})^{T}$ as weight vector, $\mathbf{\Sigma}$ defined in lemma \ref{lemme_var_estim_PP} as covariance matrix, and $\mathbf{b}=(1,\ldots,1)^{T}$ a vector of length $2N$. Thus the constant $c_{rt}$ becomes the constant $c_{pp}=\mathbf{b}^{T}(\mathbf{\Sigma})^{-1}\mathbf{b}$. \\
However, as in the case of the real-time estimator,  the computation of the optimal weights $\widehat{w}^{-}_{j}$ and $\widehat{w}^{+}_{j}$ requires the inversion of the covariance matrix $\mathbf{\Sigma}$ which is inconvenient in practice. Thus, we study the post-processing estimator with asymptotically minimum variance.


\subsubsection{Asymptotically minimum variance unbiased estimator}

\begin{Th}
\label{th_var_estim_PP_asymp}
Let $N \geq 1$ an integer, $r=\frac{\sigma^{2}_{od}}{\sigma^{2}_{gps}}$ and $\lambda=\frac{f_{od}}{f_{gps}} \in \mathds{N}^{*}$. Let $\widehat{x}^{-}_{j}$ and $\widehat{x}^{+}_{j}$, $j=1,\ldots,N$, the estimators defined in definition \ref{def_estim_PP}. Assume that $(w_{1},w_{2})$ are the weights defined in theorem \ref{th_estim_RT_asymp_recurs_weight_and_var}. Then $\widehat{x}^{\infty}_{PP}(t_{i})$ is an asymptotically minimum variance unbiased estimator of the position of the vehicle at the sampling time $t_{i}$, defined as follows:
\begin{equation}
\label{eq_estim_PP_asymp}
\widehat{x}_{PP}^{\infty}(t_{i})=\widetilde{w}_{1}\widetilde{x}_{PP}^{-}(t_{i})+\widetilde{w}_{2}\widetilde{x}_{PP}^{+}(t_{i})
\end{equation}
where

\begin{equation}
\label{eq_estim_PP_tild_moins_et_plus}
\begin{split}
&\widetilde{x}_{PP}^{-}(t_{i})=\sum_{j=1}^{N}\widetilde{w}_{j}^{-}\widehat{x}_{j}^{-}(t_{i})\ \ \ \ \ \text{and}\ \ \ \ \ \\ &\widetilde{x}_{PP}^{+}(t_{i})=\sum_{j=1}^{N}\widetilde{w}_{j}^{+}\widehat{x}_{j}^{+}(t_{i})
\end{split}
\end{equation}

with $\left\{
\begin{array}{l}
\widetilde{w}_{j}^{-}=\widetilde{w}_{j}^{+}=w_{2}w_{1}^{j-1}\ \ \text{for}\ j<N \\
\widetilde{w}_{N}^{-}=\widetilde{w}_{N}^{+}=w_{1}^{N-1}
\end{array}
\right.$.

\vspace{2ex}
The weights $(\widetilde{w}_{1},\widetilde{w}_{2})$ whose sum is equal to one, can be written as follows:

\begin{equation*}
\begin{split}
&\widetilde{w}_{1}=\frac{Var[\widetilde{x}_{PP}^{+}(t_{i})]}{Var[\widetilde{x}_{PP}^{-}(t_{i})]+Var[\widetilde{x}_{PP}^{+}(t_{i})]} \quad \text{and}\\ &\widetilde{w}_{2}=\frac{Var[\widetilde{x}_{PP}^{-}(t_{i})]}{Var[\widetilde{x}_{PP}^{-}(t_{i})]+Var[\widetilde{x}_{PP}^{+}(t_{i})]}
\end{split}
\end{equation*}

where

\begin{equation*}
\begin{split}
Var[\widetilde{x}_{PP}^{-}(t_{i})]&=\sigma_{gps}^{2}[\frac{(1-w_{1})^{2}+2w_{1}^{2N-1}(1-w_{1})}{1-w_{1}^{2}} \\
& \quad + r(d_{i}+\lambda \frac{w_{1}^{2}-w_{1}^{2N}}{1-w_{1}^{2}})] \quad \text{and} \\
 Var[\widetilde{x}_{PP}^{+}(t_{i})]&=\sigma_{gps}^{2}[\frac{(1-w_{1})^{2}+2w_{1}^{2N-1}(1-w_{1})}{1-w_{1}^{2}} \\
& \quad + r(\lambda \frac{1-w_{1}^{2N}}{1-w_{1}^{2}}-d_{i})]. \\
\end{split}
\end{equation*}

Then, we deduce the following expression for the variance of the estimator $\widehat{x}_{PP}^{\infty}$:

\begin{equation}
\label{eq_var_estim_PP_asymp}
\begin{split}
& Var[\widehat{x}_{PP}^{\infty}(t_{i})]=\frac{Var[\widetilde{x}_{PP}^{-}(t_{i})]Var[\widetilde{x}_{PP}^{+}(t_{i})]}{Var[\widetilde{x}_{PP}^{-}(t_{i})]+Var[\widetilde{x}_{PP}^{+}(t_{i})]}\\
&
\end{split}
\end{equation}
\end{Th}

As for the asymptotically minimum variance real-time estimator, when N is fixed, we obtain a truncated estimator. The asymptotic variance is obtained when N tends to infinity in the expression \refeq{eq_var_estim_PP_asymp}.

\section{Data processing and discussion}
\label{section_data_processing_and_discussion}

In this section, we present simulation and real data results and a comparison of the different estimators defined in the previous section with a classical Kalman filter. 
They have been obtained on a DELL T3400 workstation equipped with a Intel E8400 core 2 duo processor. The Kalman filter is constructed using only the odometer and GPS measurements in order to fairly compare this Kalman filter with the real-time estimators defined in the section \ref{subsection_real_time_estimator}. Thus, the state vector is only composed with one component $x_{i}$ where $x_{i}$ is the distance traveled by the vehicle at time $t_{i}$ from the initial position $x_{0}$. The dynamic equation is given by:
\begin{equation}
\label{Kalman_filter_dynamic_model}
x_{i+1}=x_{i}+(y_{od,i+1}-y_{od,i})+\varepsilon_{od,i}
\end{equation}
where $y_{od,i}$ is the odometer measurement at time $t_{i}$ and $\varepsilon_{od,i}\sim N(0,\sigma^{2}_{od})$. The measurement equation using only GPS measurement is given by:
\begin{equation}
\label{Kalman_filter_measurement_model}
y_{gps,i+1}=x_{i+1}+\varepsilon_{gps,i}
\end{equation}
where $y_{gps,i}$ is the GPS measurement at time $t_{i}$ and $\varepsilon_{gps,i}\sim N(0,\sigma^{2}_{gps})$.\\
Then the step prediction is performed as follows:
\begin{equation}
\label{Kalman prediction_step}
\left\{
\begin{array}{l}
\widehat{x}_{i+1|i}=\widehat{x}_{i|i}+(y_{od,i+1}-y_{od,i}) \\
P_{i+1|i}=P_{i|i}+\sigma^{2}_{od}
\end{array}
\right.
\end{equation}
where $\widehat{x}_{i|i}$ is the state estimate at time $t_{i}$ knowing the measures until $t_{i}$, and $P_{i|i}$ is the related covariance matrix of the estimation error (here, $P_{i|i}$ is a real number). The update step is performed as follows:
\begin{equation}
\label{Kalman update_step}
\left\{
\begin{array}{l}
K_{i+1}=P_{i+1|i}\ (P_{i+1|i}+\sigma^{2}_{gps})^{-1} \\
\widehat{x}_{i+1|i+1}=\widehat{x}_{i+1|i}+K_{i+1}(y_{gps,i+1}-\widehat{x}_{i+1|i}) \\
P_{i+1|i+1}=(1-K_{i+1})P_{i+1|i}
\end{array}
\right.
\end{equation}
The filter is initialized as follows:
\begin{equation}
\label{Kalman initialisation}
\left\{
\begin{array}{l}
\widehat{x}_{0|0}=y_{gps}(t_{0}) \\
P_{0|0}=\sigma^{2}_{gps}
\end{array}
\right.
\end{equation}
Later in the document, the Kalman filter will be denoted $\widehat{x}_{RT}^{KF}$. More details on the Kalman filter can be found in \cite{Chui1987} and \cite{Grewal2008}.

\subsection{Simulation results}
\label{subsection_simulation_results}

\begin{figure}[H]
\includegraphics[width=18cm, height=15cm]{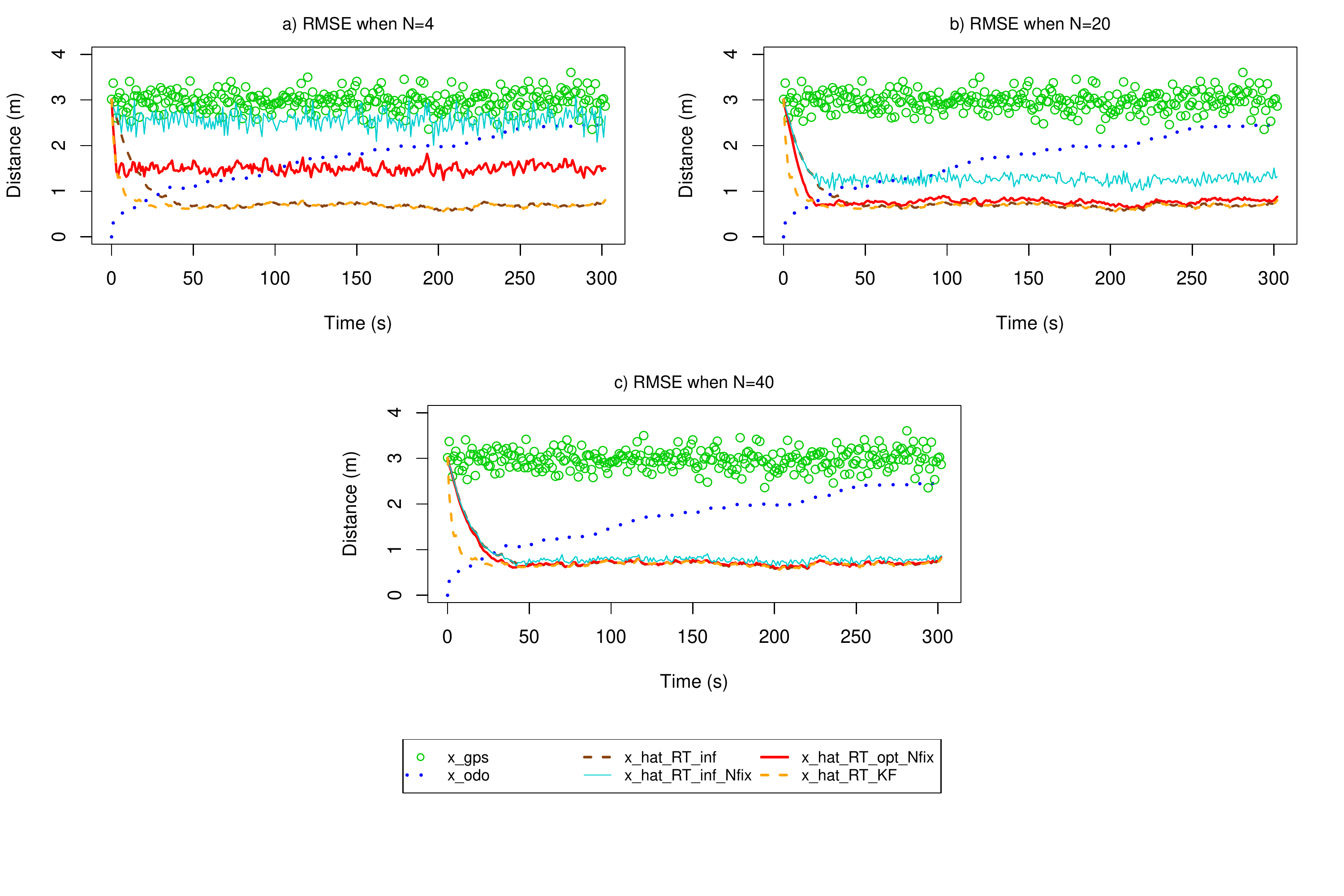}
\caption{RMSE of real-time estimators ($\widehat{x}^{\infty}_{RT}$ and $\widehat{x}^{opt}_{RT}$) and Kalman filter.}
\label{fig_rmse_estimRT_et_Kalman_100simu_freq1Hz_N4_N20_N40}
\end{figure}

\begin{figure}[H]
\includegraphics[width=18cm, height=15cm]{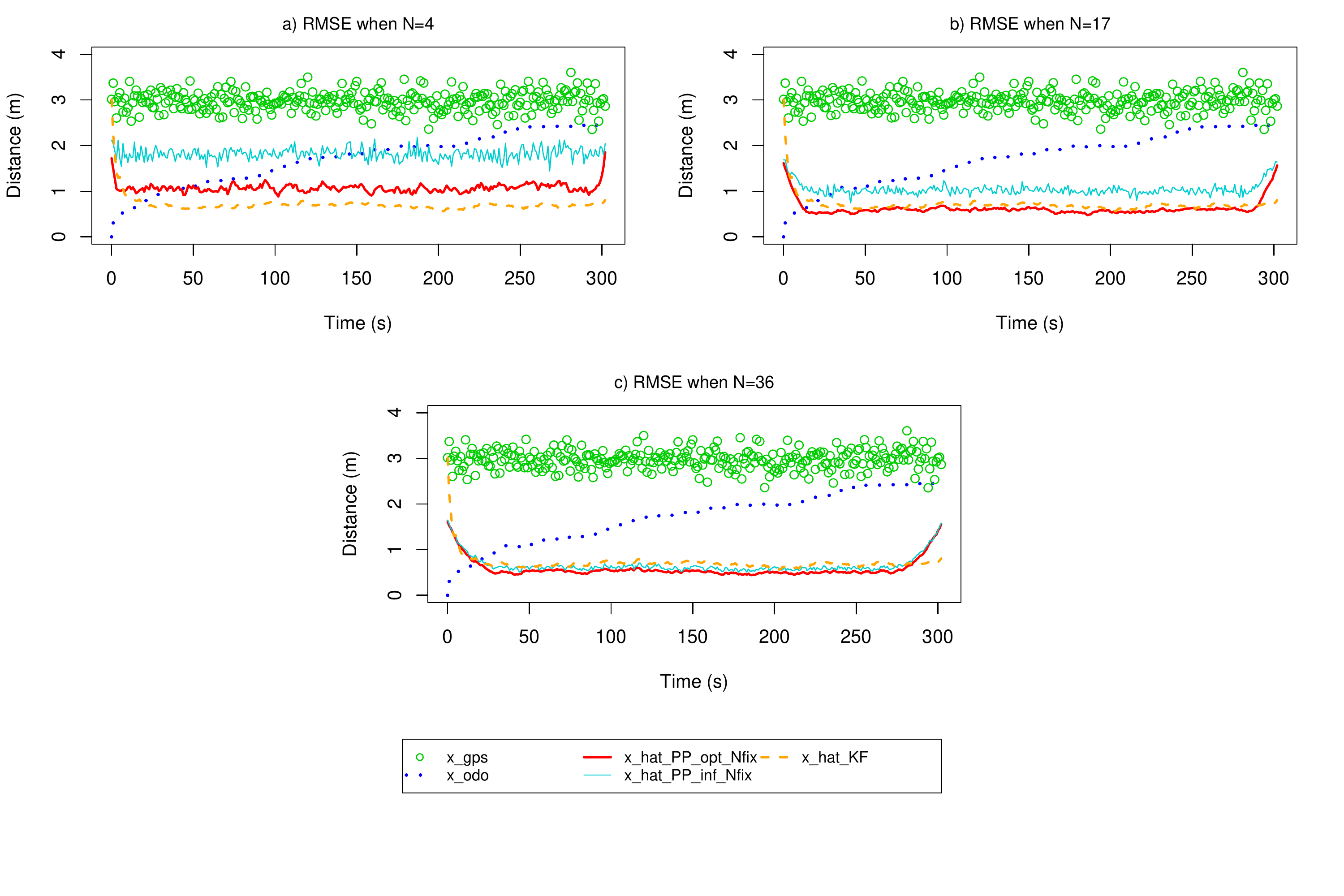}
\caption{RMSE of real-time estimators ($\widehat{x}^{\infty}_{PP}$ and $\widehat{x}^{opt}_{PP}$) and Kalman filter.}
\label{fig_rmse_estimPP_et_Kalman_100simu_complet_freq1Hz_N4_N17_N36}
\end{figure}

Given a reference vehicle trajectory length of 4000m and traveled in about 300s, the odometer and map-matched GPS data were simulated from the model \ref{eq_model_observation} with an odometer error standard deviation $\sigma_{od}$ equal to 0.05m and a GPS error standard deviation $\sigma_{gps}$ equal to 3m. These sensor simulated data represent the measurement data $y_{od}(t_{i})$ and $y_{gps}(t_{i})$ at each time $t_{i}$ used in the calculation of each estimators. We suppose that the odometer and GPS frequencies are respectively equal to 10Hz and 1HZ, so that the ratio $\lambda$ is equal to 10. 100 simulations of GPS and odometer measurements were generated, each simulation involved generating a new set of sensor data of the reference distance traveled and computing the estimated location at each time $t_{i}$ with each estimators defined in the previous section (real-time estimators and post-processing estimators) and with a Kalman filter. For each estimator, the Root Mean Square Error (RMSE) for all 100 simulations was computed every second (i.e. at each time $t_{i}$ for which a GPS measurement is available). \\
The RMSE is a good measure of the accuracy of an estimator and has the advantage of being expressed in the same units as the quantity being estimated (i.e. in meters). The RMSE of an estimator $\widehat{X}$ of a vector $X$ is defined as follows:
\begin{equation}
\label{def_rmse}
\begin{split}
RMSE(\widehat{X})&=\sqrt{MSE(\widehat{X})}\\
&=\sqrt{E((\widehat{X}-X)^{T}(\widehat{X}-X))}\\
&=\sqrt{\frac{1}{n} \sum_{i=1}^{n}(\widehat{X}_{i}-X_{i})^{2}}
\end{split}
\end{equation}
where $\widehat{X}_{i}$ (resp. $X_{i}$) is the i-th component of the vector $\widehat{X}$ (resp. $X$). There are other types of errors (e.g. mean absolute error, geometric average error), but the choice of the MSE is justified by its interpretation in terms of bias and variance:
\begin{equation}
\label{def_rmse_bias_variance}
MSE(\widehat{X})=[Bias(\widehat{X})]^{2} + Var(\widehat{X})
\end{equation}
where $Bias(\widehat{X})=E[\widehat{X}]-X$. Thus, the best estimator between two unbiased estimators is the one that has the smallest variance, and an unbiased estimator of minimum variance is generally regarded as the best estimator possible. It is moreover well as real-time estimator $\widehat{x}^{opt}_{RT}$ and post-processing estimator $\widehat{x}^{opt}_{PP}$ were built. \\
\\
Figures \ref{fig_rmse_estimRT_et_Kalman_100simu_freq1Hz_N4_N20_N40} and \ref{fig_rmse_estimPP_et_Kalman_100simu_complet_freq1Hz_N4_N17_N36} contain the RMSE of the vehicle location obtained with each estimator. In these two figures, the RMSE of the simulated sensors are represented by green circles for GPS and blue dots for the odometer. Each estimator (real-time and post-processing) is compared with the Kalman filter $\widehat{x}_{RT}^{KF}$ defined in the beginning of the section \ref{section_data_processing_and_discussion} (denoted x\_hat\_RT\_KF in Figure \ref{fig_rmse_estimRT_et_Kalman_100simu_freq1Hz_N4_N20_N40} and \ref{fig_rmse_estimPP_et_Kalman_100simu_complet_freq1Hz_N4_N17_N36}) and represented by orange dashed line. Figure \ref{fig_rmse_estimRT_et_Kalman_100simu_freq1Hz_N4_N20_N40} presents the comparaison between the RMSE of the vehicle location obtained with the real-time estimators defined in section \ref{subsection_real_time_estimator} and the Kalman filter $\widehat{x}_{RT}^{KF}$. The chocolate dashed line represent the asymptotically minimum variance estimator $\widehat{x}_{RT}^{\infty}$ (denoted x\_hat\_RT\_inf in Figure \ref{fig_rmse_estimRT_et_Kalman_100simu_freq1Hz_N4_N20_N40}) defined recursively in theorem \ref{th_estim_RT_asymp_recurs_weight_and_var} and initialized with $\widehat{x}^{\infty}_{RT}(t_{0})=y_{gps}(t_{0})$. The turquoise line represented the truncated estimator $\widehat{x}_{RT}^{\infty}$ with a fixed N defined in theorem \ref{th_var_estim_RT_asymp_nonrecurs} (and denoted x\_hat\_RT\_inf\_Nfix in Figure \ref{fig_rmse_estimRT_et_Kalman_100simu_freq1Hz_N4_N20_N40}) and the red line represent the minimum variance estimator $\widehat{x}_{RT}^{opt}$ for a fixed N defined in theorem \ref{th_estim_RT_opt} (and denoted x\_hat\_RT\_opt\_Nfix in Figure \ref{fig_rmse_estimRT_et_Kalman_100simu_freq1Hz_N4_N20_N40}). These two estimators depending on N were computed for three different values of N:
\begin{itemize}
  \item N=4 is a small value chosen at random;
  \item N=20 is the threshold above which the difference between the standard deviation of the minimum variance estimator $\sqrt{Var[\widehat{x}_{RT}^{opt}(t_{i})]}$ defined in theorem \ref{th_estim_RT_opt} (with $d_{i}=0$ since the time step is 1s and the GPS frequency is 1Hz) and the square root of the asymptotic variance of $\widehat{x}_{RT}^{\infty}$ defined in \refeq{eq_var_estim_RT_asymp_recurs} is less than 0.1m;
  \item N=40 is the threshold above which the difference between the standard deviation of the truncated asymptotically minimum variance estimator $\sqrt{Var[\widehat{x}_{RT}^{\infty}(t_{i})]}$ with N fixed defined in theorem \ref{th_var_estim_RT_asymp_nonrecurs} (with $d_{i}=0$) and the square root of the asymptotic variance of $\widehat{x}_{RT}^{\infty}$ defined in \refeq{eq_var_estim_RT_asymp_recurs} is less than 0.1m.
\end{itemize}
Figure \ref{fig_rmse_estimRT_et_Kalman_100simu_freq1Hz_N4_N20_N40} shows that the Kalman filter is the best estimator of the vehicle location, mainly at each time $t_{i}$ of the beginning of the path, but after around 50s the RMSE curve of the Kalman filter and that of the asymptotically minimum variance estimator $\widehat{x}_{RT}^{\infty}$ are merged. Similarly, when N=20, the RMSE curve of the minimum variance estimator $\sqrt{Var[\widehat{x}_{RT}^{opt}(t_{i})]}$ is approximately merged with the RMSE curve of the Kalman filter, and it is the same for the truncated version of the estimator $\widehat{x}_{RT}^{\infty}$ when N=40. The average and maximum RMSE of each estimator represented in Figure \ref{fig_rmse_estimRT_et_Kalman_100simu_freq1Hz_N4_N20_N40} are given in Table \ref{tab_rmse_mean_and_max_estim_RT}. These values confirm the results described in Figure \ref{fig_rmse_estimRT_et_Kalman_100simu_freq1Hz_N4_N20_N40}. The asymptotic standard deviation achieved by all estimators and equal to $\sqrt{Var[\widehat{x}_{RT}^{\infty}(t_{i})]}$ when $t_{i} \rightarrow \infty$ can be calculated with the formula given in \refeq{eq_var_estim_RT_asymp_recurs}. We then obtained an optimal standard deviation equal to 0.68m which corresponds approximatively to the RMSE obtained with each estimator after 50s when N is sufficiently large.

\begin{table}[H]
  \begin{tabular}{|>{\centering}m{6.5cm}|>{\centering}m{3.5cm}|m{3.5cm}<{\centering}|}
  \hline
     & Mean RMSE (m) & Max RMSE (m)  \\ \hline\hline
  Odometer                                             & 1.72 & 2.55  \\ \hline
  GPS                                                  & 2.97 & 3.60  \\ \hline
  $\widehat{x}_{RT}^{\infty}$                          & 0.79 & 3.01  \\ \hline
  $\widehat{x}_{RT}^{\infty}$ with N fixed (N=4)       & 2.57 & 3.08  \\ \hline
  $\widehat{x}_{RT}^{\infty}$ with N fixed (N=20)      & 1.31 & 3.01  \\ \hline
  $\widehat{x}_{RT}^{\infty}$ with N fixed (N=40)      & 0.86 & 3.01  \\ \hline
  $\widehat{x}_{RT}^{opt}$ with N fixed (N=4)          & 1.51 & 3.01  \\ \hline
  $\widehat{x}_{RT}^{opt}$ with N fixed (N=20)         & 0.83 & 3.01  \\ \hline
  $\widehat{x}_{RT}^{opt}$ with N fixed (N=40)         & 0.78 & 3.01  \\ \hline
  $\widehat{x}_{RT}^{KF}$                              & 0.71 & 3.01  \\ \hline
  \end{tabular}
\caption{Mean and maximum RMSE of the real-time estimators compared to the Kalman filter.}
\label{tab_rmse_mean_and_max_estim_RT}
\end{table}

Figure \ref{fig_rmse_estimPP_et_Kalman_100simu_complet_freq1Hz_N4_N17_N36} and Table \ref{tab_rmse_mean_and_max_estim_PP} are similar to Figure \ref{fig_rmse_estimRT_et_Kalman_100simu_freq1Hz_N4_N20_N40} and Table \ref{tab_rmse_mean_and_max_estim_RT} but with a comparison between the Kalman filter and the post-processing estimators defined in section \ref{subsection_post_processing_estimator}. Thus, the turquoise line represents the truncated estimator $\widehat{x}_{PP}^{\infty}$ with a fixed N defined in theorem \ref{th_var_estim_PP_asymp} (and denoted x\_hat\_PP\_inf\_Nfix in Figure \ref{fig_rmse_estimPP_et_Kalman_100simu_complet_freq1Hz_N4_N17_N36}) and the red line represents the minimum variance estimator $\widehat{x}_{PP}^{opt}$ for a fixed N defined after the lemma \ref{lemme_var_estim_PP} (and denoted x\_hat\_PP\_opt\_Nfix in Figure \ref{fig_rmse_estimPP_et_Kalman_100simu_complet_freq1Hz_N4_N17_N36}). These two estimators depending on N were computed for N=4 as in the real-time case and also for the two following values:
\begin{itemize}
  \item N=17 is the threshold above which the difference between the standard deviation of the minimum variance estimator $\sqrt{Var[\widehat{x}_{PP}^{opt}(t_{i})]}$ (with $d_{i}=0$) and the square root of the asymptotic variance of $\widehat{x}_{PP}^{\infty}$ obtained in \refeq{eq_var_estim_PP_asymp} when N tend to infinity, is less than 0.1m;
  \item N=36 is the threshold above which the difference between the standard deviation of the truncated asymptotically minimum variance estimator $\sqrt{Var[\widehat{x}_{PP}^{\infty}(t_{i})]}$ with N fixed defined in theorem \ref{th_var_estim_PP_asymp} (with $d_{i}=0$) and the square root of the asymptotic variance of $\widehat{x}_{PP}^{\infty}$ obtained in \refeq{eq_var_estim_PP_asymp} when N tend to infinity, is less than 0.1m.
\end{itemize}
Figure \ref{fig_rmse_estimPP_et_Kalman_100simu_complet_freq1Hz_N4_N17_N36} and Table \ref{tab_rmse_mean_and_max_estim_PP} show that when we use measurements obtained after time $t_{i}$ (post-processing case), the accuracy of the estimate of the position of the vehicle is improved and is better than using the Kalman filter except at the end of the path where a side effect appears. The asymptotic standard deviation achieved by all estimators when N is sufficiently large, except on the boundaries, and corresponding to $\sqrt{Var[\widehat{x}_{PP}^{\infty}(t_{i})]}$ when $N \rightarrow \infty$ in \refeq{eq_var_estim_PP_asymp}, is equal to 0.49m.

\begin{table}[H]
  \begin{tabular}{|>{\centering}m{6.5cm}|>{\centering}m{3.5cm}|m{3.5cm}<{\centering}|}
  \hline
     & Mean RMSE (m) & Max RMSE (m)  \\ \hline\hline
  Odometer                                             & 1.72 & 2.55  \\ \hline
  GPS                                                  & 2.97 & 3.60  \\ \hline
  $\widehat{x}_{PP}^{opt}$ with N fixed (N=4)          & 1.07 & 1.86  \\ \hline
  $\widehat{x}_{PP}^{opt}$ with N fixed (N=17)         & 0.62 & 1.62  \\ \hline
  $\widehat{x}_{PP}^{opt}$ with N fixed (N=36)         & 0.59 & 1.62  \\ \hline
  $\widehat{x}_{PP}^{\infty}$ with N fixed (N=4)       & 1.82 & 2.18  \\ \hline
  $\widehat{x}_{PP}^{\infty}$ with N fixed (N=17)      & 1.04 & 1.69  \\ \hline
  $\widehat{x}_{PP}^{\infty}$ with N fixed (N=36)      & 0.66 & 1.65  \\ \hline
  $\widehat{x}_{RT}^{KF}$                              & 0.71 & 3.01  \\ \hline
  \end{tabular}
\caption{Mean and maximum RMSE of the post-processing estimators compared to the Kalman filter.}
\label{tab_rmse_mean_and_max_estim_PP}
\end{table}

\subsection{Real data results}
\label{subsection_real_data_results}

In this section, real data collected from a trip provided on test tracks at Versailles-Satory (France) were used. The trip length was around 4000m and a travel time of 300s. The odometer measurements have been collected on CAN (Controller Area Network) bus of the vehicle and have been provided at a 10Hz sampling frequency. Two GPS were also located on the roof of the vehicle: A GlobalSat BR-355 GPS receiver (with SIRF Star III) and a Thales Sagitta RTK-GPS receiver. The BR-355 GPS provides position measurements at a 1Hz sampling frequency with a 10m accuracy and the RTK-GPS (Real-Time Kinematic Global Positioning System) provides position measurements at a 10Hz sampling frequency with a centimeter accuracy. Thus, the RTK-GPS measurements were used as the "true" locations of the vehicle and were considered as the reference trajectory. A simple map-matching algorithm was used in order to project the GPS measurements on the road, and the position measurements from the odometer and the two GPS were synchronized in time. According to the accuracy of the sensors, we assume that the standard deviations of the errors of the odometer $\sigma_{od}$ and the GPS $\sigma_{gps}$ are respectively equal to 0.03m and 3m. The ratio $\lambda$ between the odometer and GPS sampling frequencies is equal to 10 as in the previous section. \\
Tables \ref{tab_rmse_and_CP_estim_RT_and_PP}.a and \ref{tab_rmse_and_CP_estim_RT_and_PP}.b contain the RMSE of the vehicle location of each estimator on the complete trip, and the total computation time of the estimated positions with each estimator at each time $t_{i}$ with a sampling frequency of 10Hz. The Kalman filter is compared with the real-time estimators in Table \ref{tab_rmse_and_CP_estim_RT_and_PP}.a. For comparaison, the values of N are the same as in the previous section with simulated data. Contrary to previous results obtained with simulated data, Table \ref{tab_rmse_and_CP_estim_RT_and_PP}.a shows that the minimum variance estimator $\widehat{x}_{RT}^{opt}$ and the truncated version of the estimator $\widehat{x}_{RT}^{\infty}$ with a fixed N are better than the Kalman filter and the asymptotically minimum variance estimator $\widehat{x}_{RT}^{\infty}$. However the results show that the GPS is more accuracy (RMSE=3.07m) than all the real-time estimators except the minimum variance estimator $\widehat{x}_{RT}^{opt}$ with N=4 (RMSE=2.59m). Furthermore, increasing the interval smoothing (i.e. increase the value of N) does not improve the accuracy of the real-time estimators that depend on N. Table \ref{tab_rmse_and_CP_estim_RT_and_PP}.b shows that all the post-processing estimators are better than the real-time estimators and are more accuracy than the GPS. However the computational time of the post-processing estimators are bigger than the real-time estimators mainly when N is large.

\begin{table}[H]
\begin{tabular}{>{\centering}m{8cm} m{8cm}<{\centering}}
a) Real-time estimators & b) Post-processing estimators \\

  \begin{tabular}{|m{3cm}|>{\centering}m{1.5cm}|m{2cm}<{\centering}|}
  \hline
     & RMSE (m) & Computing time (s)  \\ \hline\hline
  Odometer                                & 20.59 &  -  \\ \hline
  GPS                                     & 3.07 &  -  \\ \hline
  $\widehat{x}_{RT}^{KF}$                 & 4.37 & 0.06 \\ \hline \hline
  $\widehat{x}_{RT}^{\infty}$             & 4.50 & 0.03 \\ \hline
  $\widehat{x}_{RT}^{\infty}$ (N=4)       & 3.21 & 0.26 \\ \hline
  $\widehat{x}_{RT}^{\infty}$ (N=20)      & 3.91 & 1.01 \\ \hline
  $\widehat{x}_{RT}^{\infty}$ (N=40)      & 3.93 & 1.94 \\ \hline \hline
  $\widehat{x}_{RT}^{opt}$ (N=4)          & 2.59 & 0.28 \\ \hline
  $\widehat{x}_{RT}^{opt}$ (N=20)         & 3.08 & 1.10 \\ \hline
  $\widehat{x}_{RT}^{opt}$ (N=40)         & 3.52 & 2.09 \\ \hline
  \end{tabular}
&
  \begin{tabular}{|m{3cm}|>{\centering}m{1.5cm}|m{2cm}<{\centering}|}
  \hline
     & RMSE (m) & Computing time (s)  \\ \hline\hline
  Odometer                                & 20.59 &  -  \\ \hline
  GPS                                     & 3.07 &  -  \\ \hline \hline
  $\widehat{x}_{PP}^{opt}$ (N=4)          & 2.28 & 0.54 \\ \hline
  $\widehat{x}_{PP}^{opt}$ (N=17)         & 1.54 & 1.94 \\ \hline
  $\widehat{x}_{PP}^{opt}$ (N=36)         & 1.54 & 3.97 \\ \hline \hline
  $\widehat{x}_{PP}^{\infty}$ (N=4)       & 2.42 & 0.49 \\ \hline
  $\widehat{x}_{PP}^{\infty}$ (N=17)      & 1.67 & 1.78 \\ \hline
  $\widehat{x}_{PP}^{\infty}$ (N=36)      & 2.08 & 3.72 \\ \hline
  \end{tabular} \\
\end{tabular}
\caption{RMSE and computing time of the real-time and post-processing estimators compared to the Kalman filter.}
\label{tab_rmse_and_CP_estim_RT_and_PP}
\end{table}

\subsection{Discussion}
\label{subsection_discussion}

The asymptotically minimum variance estimator $\widehat{x}_{RT}^{\infty}$ is similar to the Kalman filter $\widehat{x}_{RT}^{KF}$ with a fixed gain $K$ which is optimal when the estimation time $t_{i}$ tends to infinity. Indeed, the weights $(w_{1},w_{2})$ defined in definition \ref{def_estim_RT_asymp_recurs} and theorem \ref{th_estim_RT_asymp_recurs_weight_and_var} are fixed which saves computation time (twice as fast as the Kalman filter with the real data). Furthermore, the construction of $\widehat{x}_{RT}^{\infty}$ provides a simple expression of the asymptotic variance (equation \refeq{eq_var_estim_RT_asymp_recurs}). However, the optimality of these weights only at infinity implies a poor accuracy of the estimator $\widehat{x}_{RT}^{\infty}$ at the beginning of the trip, even if its convergence to the optimal estimator is relatively fast (50s with the simulated data). Thus, when the vehicle distance traveled to be estimated is quite long in time, the estimator $\widehat{x}_{RT}^{\infty}$ can be more effective.\\
By definition, the estimator $\widehat{x}_{RT}^{opt}$ is the optimal estimator at each time $t_{i}$ for a N fixed and the Kalman filter $\widehat{x}_{RT}^{KF}$ is the optimal estimator using all measurements available up to $t_{i}$. Therefore, at a given time $t_{i}$, the Kalman filter $\widehat{x}_{RT}^{KF}(t_{i})$ is similar to the estimator $\widehat{x}_{RT}^{opt}(t_{i})$ with $N = \lfloor\frac{i}{\lambda}\rfloor+1$ (corresponding to use all measurements up to time $t_{i}$). The Kalman filter $\widehat{x}_{RT}^{KF}(t_{i})$ then corresponds to $\widehat{x}_{RT}^{opt}$ with a non-fixed interval filtering which increases over time. However, contrary to the simulation results, the real data results have shown that increasing the size of the interval filtering, by increasing N, does not improve the accuracy of the estimator $\widehat{x}_{RT}^{opt}$. Indeed, adding too much information to estimate the position at a sampling time $t_{i}$ can bias the estimation. Measurements obtained at times close to $t_{i}$ are supposed to contain the most accurate information to estimate the position at time $t_{i}$ unless such measures are very noisy. In the case of noisy measurements around $t_{i}$, it is better to increase the interval filtering (or the interval smoothing in the post-processing case) even if the computational time increase. Thus, the best estimator could be an estimator with a variable interval filtering that would be optimal at each time $t_{i}$ even if the computation time of such an estimator would certainly be large. However, it is important to note that Tables \ref{tab_rmse_and_CP_estim_RT_and_PP}.a and \ref{tab_rmse_and_CP_estim_RT_and_PP}.b give the computational time to estimate the whole trip, i.e. a vector of size $n$ containing the estimated distance traveled at each time $t_{i}$, which benefits recursive estimators as the Kalman filter. But if we want only one estimated distance traveled at a given time $t_{i}$, a recursive expression requires to compute all the estimated distance up to $t_{i}$, which can significantly increase the computational time when $t_{i}$ is large. For example, with the real data used in section \ref{subsection_real_data_results}, the computational time of the estimated position at time $t_{i}=300s$ is respectively 6.30ms for the Kalman filter and 2.45ms for the estimator  $\widehat{x}_{RT}^{opt}$ with $N=20$. Thus, in some cases, the estimator $\widehat{x}_{RT}^{opt}(t_{i})$ is faster to compute, even if the computational time depends on the size of the interval filtering or smoothing. \\
Finally, the estimator $\widehat{x}_{RT}^{opt}$ (resp. $\widehat{x}_{PP}^{opt}$) is accurate but requires the inversion of a matrix of size $N\times N$ (resp. $2N\times 2N$) which can be a disadvantage when N is large. In such cases, the truncated version of $\widehat{x}_{RT}^{\infty}$ (resp. $\widehat{x}_{PP}^{\infty}$) with a fixed N can be a good alternative. Indeed, this estimator is faster to compute than $\widehat{x}_{RT}^{opt}$ (resp. $\widehat{x}_{PP}^{opt}$) for the same N and converge to the optimal estimator when N tends to infinity, even if it is on average less accuracy.

\section{Conclusion}
Some real-time and post-processing estimators of the distance traveled by a vehicle on a road segment was developed and compared to a classic Kalman filter. These estimators were based on a simple statistical error model of the odometer and the GPS which makes the model generalizable to other applications. Firstly, a recursive asymptotically minimum variance filter, similar to the Kalman filter with a fixed gain K which is optimal when time tends to infinity, was developed. This estimator is two times faster to compute than the Kalman filter and converges quickly to the optimal estimator (error less than 1m after 50s with simulated data). Then, two more flexible filters was developed using only measurements included in a moving fixed-interval: an optimal filter that requires the inversion of the covariance matrix, and a truncated version of the asymptotically minimum variance filter that is less accurate but have a simpler expression of weights and variance. Real-data results have shown the interest of using moving fixed-interval filter instead of recursive filter such as the Kalman filter: the error can be averaged less than 3m with a good choice of filtering window size. Finally, two moving fixed-interval smoothers derived from the two previous filters was also developed for post-processing cases. These smoothers estimate the vehicle position at a time $t$ by using the measurement over a specified window around $t$. Taking into account both past and future observations allows to achieve better estimation accuracy (error less than 2m with real data). \\
In future work, the robustness of the estimators presented in this paper will be tested. Furthermore, this work have shown the interest of using moving fixed-interval filter/smoother but the choice of the best filtering/smoothing windows size for the whole trip is difficult. The development of a nonfixed-interval filter/smoother with an optimal windows size at each time $t$ could be a good alternative even if it would certainly be at the expense of a higher computation time.

\section{Appendix A: Nomenclature}

\nomenclature[a]{$x(t_{i})$}{vehicle position at time $t_{i}$, $t_{i}\in [0,T]$}%
\nomenclature[b]{$y_{od}(t_{i}), y_{gps}(t_{i})$}{odometer and map-matched GPS measurements at time $t_{i}$, $t_{i}\in [0,T]$}%
\nomenclature[c]{$\sigma^{2}_{od}, \sigma^{2}_{gps}$}{variances of the odometer and the GPS}%
\nomenclature[d]{$r$}{ratio between the variance of the odometer and the variance of the GPS}%
\nomenclature[e]{$\lambda$}{ratio between the sampling frequency of the odometer and the sampling frequency of the GPS}%
\nomenclature[f]{$d_{i}$}{Number of odometer measurements between the time $t_{i}$ and the first time of a GPS measurement before $t_{i}$}%
\nomenclature[g]{$N$}{Number of GPS measurements used in the computation of the fixed-interval filter/smoother}%
\nomenclature[h]{$\widehat{x}^{KF}_{RT}(t_{i})$}{Kalman filter estimator of the vehicle position at time $t_{i}$}%
\nomenclature[i]{$\widehat{x}^{\infty}_{RT}(t_{i})$}{Real-time unbiased estimator of the vehicle position at time $t_{i}$ with asymptotically minimum variance}%
\nomenclature[j]{$\widehat{x}^{opt}_{RT}(t_{i})$}{Real-time unbiased estimator of the vehicle position at time $t_{i}$ with minimum variance for a fixed N}%
\nomenclature[k]{$\widehat{x}^{\infty}_{PP}(t_{i})$}{Post-processing unbiased estimator of the vehicle position at time $t_{i}$ with asymptotically minimum variance}%
\nomenclature[l]{$\widehat{x}^{opt}_{PP}(t_{i})$}{Post-processing unbiased estimator of the vehicle position at time $t_{i}$ with minimum variance for a fixed N}%

\printnomenclature[7em] 

\section{Appendix B: Proofs of Theorems}

\begin{pfof}{Theorem \ref{th_estim_RT_asymp_recurs_weight_and_var}}
~\\
\textbf{Unbiased:} Since $\varepsilon_{od,i}$ and $\varepsilon_{gps,j}$ are centered errors, it is easy to prove that $E[\widehat{x}^{\infty}_{RT}(t_{i})-x(t_{i})]=0$, i.e. the estimator is unbiased. \\
\textbf{Convergence of variance:} Considering only the sampling time where we have both odometer and GPS measurements, the two equations defined in \refeq{eq_estim_RT_asymp_recurs} can be written as follows:
$\widehat{x}^{\infty}_{RT}(t_{\lambda j})=
w_{1}\ [\widehat{x}^{\infty}_{RT}(t_{\lambda (j-1)})+\sum_{k=\lambda(j-1)}^{\lambda j-1}(y_{od}(t_{k+1})-y_{od}(t_{k}))] + w_{2}\ y_{gps}(t_{\lambda j}), \ \ j=1,\ldots,m$.
Thus, the variance of the estimator can be represented as an arithmetico-geometric sequence: $Var[\widehat{x}^{\infty}_{RT}(t_{\lambda j})]=w^{2}_{1}\ Var[\widehat{x}^{\infty}_{RT}(t_{\lambda (j-1)})] +w^{2}_{1}\ \lambda\ \sigma^{2}_{od}+w^{2}_{2}\ \sigma^{2}_{gps},\ \ j=1,...,m$.
Since $|w^{2}_{1}|<1$, the sequence converges and its limit is
$\sigma^{2}_{od}\ \frac{\lambda\ w^{2}_{1}+\frac{1}{r}w^{2}_{2}}{1-w^{2}_{1}}$ where $r=\frac{\sigma^{2}_{od}}{\sigma^{2}_{gps}}$.\\
\textbf{Calculation of asymptotically optimal weights:} The sum of weights is equal to one. Thus, the asymptotic variance can be written as a function of one variable defined on $[0,1]$ as follows:
$\varphi(w_{1})=\sigma^{2}_{od}\ \frac{\lambda\,w_{1}^{2}+\frac{1}{r}(1-w_{1})^{2}}{1-w_{1}^{2}}$. Since $\varphi$ is convex on $[0,1]$, $\varphi$ has a global minimum which satisfies the quadratic equation $\varphi'(w_{1})=0$. This equation has only one solution on $[0,1]$: $w_{1}=\frac{\lambda\,r+2-\sqrt{\lambda\,r(\lambda\,r+4)}}{2}$.
\end{pfof}

~\\
\begin{pfof}{Theorem \ref{th_var_estim_RT_asymp_nonrecurs}}
~\\
The matrix form \ref{eq_var_matrix_estim_RT_asymp_nonrecurs} of the variance of $\widehat{x}^{\infty}_{RT}$ is derived from the definition \ref{eq_estim_RT_asymp_nonrecurs}. Furthermore, for a given sampling time $t_{i}$ and for $j=1,\ldots,N$, $Var[\widehat{x}^{-}_{j}(t_{i})]=\sigma^{2}_{gps}+(d_{i}+(j-1)\lambda)\,\sigma^{2}_{od}$ according to the definition of the estimators $\widehat{x}^{-}_{j}$ given in \ref{eq_estim_moins}, and $Cov(\widehat{x}^{-}_{j}(t_{i}),\widehat{x}^{-}_{j'}(t_{i}))
= Var[\sum_{k=g^{-}_{i}(j)+1}^{i}\varepsilon_{od,k}]
= (d_{i}+(j-1)\lambda)\,\sigma^{2}_{od}$ by independence of $\varepsilon_{od,i}$, which proves the expression \ref{eq_covariance_matrix_sigma_moins} of the covariance matrix $\mathbf{\Sigma^{-}}$. Now, it remains to show the linear expression \ref{eq_var_linear_estim_RT_asymp_nonrecurs} of the variance of $\widehat{x}^{\infty}_{RT}$. We have shown that: \\
$Var[\widehat{x}^{\infty}_{RT}(t_{i})] = (\mathbf{\widetilde{w}^{-}})^{T}\, \mathbf{\Sigma^{-}} \mathbf{\widetilde{w}^{-}}
= \sigma_{gps}^{2} [(\mathbf{\widetilde{w}^{-}})^{T}\,\mathbf{ \widetilde{w}^{-}} + r\, (\mathbf{\widetilde{w}^{-}})^{T}\, \mathbf{A_{N}}(d_{i})\, \mathbf{\widetilde{w}^{-}}]$ where \\
$(\mathbf{\widetilde{w}^{-}})^{T}\, \mathbf{\widetilde{w}^{-}}
= \sum_{j=1}^{N} (\widetilde{w}_{j}^{-})^{2}
= w_{2}^{2} \sum_{j=1}^{N-1}(w_{1}^{2})^{j-1}+(w_{1}^{2})^{N-1}
= (1-w_{1})^{2} \frac{1-(w_{1}^{2})^{N-1}}{1-w_{1}^{2}} + (w_{1}^{2})^{N-1}
= \frac{(1-w_{1})^{2}+2w_{1}^{2N-1}(1-w_{1})}{1-w_{1}^{2}}$ \\
and $(\mathbf{\widetilde{w}^{-}})^{T}\, \mathbf{A_{N}}(d_{i})\, \mathbf{\widetilde{w}^{-}}
= (\mathbf{\widetilde{w}^{-}})^{T}\, (d_{i} \mathbf{H_{N}} + \lambda \mathbf{\Delta_{N})}\, \mathbf{\widetilde{w}^{-}}$ with
$\mathbf{H_{N}}=\begin{bmatrix}
1 & \cdots & 1 \\
\vdots & \ddots & \vdots \\
1 & \cdots & 1
\end{bmatrix}$
and  \\
$\mathbf{\Delta_{N}}=\begin{bmatrix}
0 & \cdots  & \cdots  & \cdots & 0 \\
\vdots  & 1 & \cdots  & \cdots & 1 \\
\vdots  & \vdots & 2& \cdots & 2\\
\vdots & \vdots & \vdots & \ddots & \vdots \\
0 & 1 & 2 & \cdots & N-1
\end{bmatrix}$. \\
We prove easily that $(\mathbf{\widetilde{w}^{-}})^{T}\, \mathbf{H_{N}}\, \mathbf{\widetilde{w}^{-}}=(\sum_{j=1}^{N} (\widetilde{w}_{j}^{-}))^{2}=1$ and if we decompose $\mathbf{\Delta_{N}}$ as follows:


\begin{equation*}
\begin{split}
 \mathbf{\Delta_{N}}  &=
              \begin{bmatrix}
                  0 & \cdots  & \cdots  & 0 \\
                  \vdots & 1 & \cdots & 1 \\
                  \vdots & \vdots & \ddots & \vdots \\
                  0 & 1 & \cdots & 1
              \end{bmatrix}
              +
              \begin{bmatrix}
                  0 & \cdots  & \cdots  & \cdots & 0 \\
                  \vdots & 0 & \cdots & \cdots & 0 \\
                  \vdots & \vdots & 1 & \cdots & 1 \\
                  \vdots & \vdots & \vdots & \ddots & \vdots \\
                  0 & 0 & 1 & \cdots & 1
              \end{bmatrix}\\
               &\quad + \cdots
              +
              \begin{bmatrix}
                  0 & \cdots  & \cdots  & 0 \\
                  \vdots & \ddots & \vdots & \vdots \\
                  0 & \cdots  & \cdots  & 0 \\
                  0 & \cdots & 0 & 1
              \end{bmatrix}\\
               &=  \mathbf{C_{1}} + \mathbf{C_{2}} + \ldots + \mathbf{C_{N-1}}
\end{split}
\end{equation*}

we obtain that: \\
$(\mathbf{\widetilde{w}^{-}})^{T}\, \mathbf{\Delta_{N}}\, \mathbf{\widetilde{w}^{-}}
= \sum_{k=1}^{N-1} ((\mathbf{\widetilde{w}^{-}})^{T}\, \mathbf{C_{k}}\, \mathbf{\widetilde{w}^{-}})
= \sum_{k=1}^{N-1} (\sum_{j=k+1}^{N} (\widetilde{w}_{j}^{-}))^{2}
= \sum_{k=2}^{N} (\sum_{j=k}^{N} (\widetilde{w}_{j}^{-}))^{2}$ where
$\sum_{j=k}^{N} \widetilde{w}_{j}^{-}
= \sum_{j=k}^{N-1} w_{2} w_{1}^{k-1}\frac{1-w_{1}^{N-k}}{1-w_{1}}+w_{1}^{N-1}
= w_{1}^{k-1} (1-w_{1}^{N-k})+w_{1}^{N-1}
= w_{1}^{k-1}$.\\
Then we deduce that
$(\mathbf{\widetilde{w}^{-}})^{T}\, \mathbf{\Delta_{N}}\, \mathbf{\widetilde{w}^{-}}
= \sum_{k=2}^{N} w_{1}^{2(k-1)}
= \frac{w_{1}^{2}-w_{1}^{2N}}{1-w_{1}^{2}}$ which completes the proof of the expression \ref{eq_var_linear_estim_RT_asymp_nonrecurs}.
\end{pfof}

~\\
\begin{pfof}{Theorem \ref{th_estim_RT_opt}}
~\\
We search the optimal weights $\mathbf{\widehat{w}^{-}}=(\widehat{w}^{-}_{1},\ldots,\widehat{w}^{-}_{N})^{T}$ that minimize the variance of the estimator under the condition that the sum of weights is equal to 1, i.e. the following constraint optimization problem:
\begin{equation*}
\left\{
\begin{array}{l}
\displaystyle{\min_{\mathbf{w}}\mathbf{w}^{T}\mathbf{\Sigma^{-}}\,\mathbf{w}}
\\
\text{subject to}\ \ \mathbf{w}^{T}\mathbf{b}=1
\end{array}
\right.
\end{equation*}
This is an optimization problem of a quadratic function with an equality constraint. The Lagrange multiplier method is used to solve this optimization problem. The Lagrangian function is $L(\mathbf{w},\lambda)=\mathbf{w}^{T}\mathbf{\Sigma^{-}}\,\mathbf{w} + \lambda(1-\mathbf{w}^{T}\mathbf{b})$ where $\lambda$ is the Lagrange multiplier. The first optimality condition is:
\begin{equation}
\label{eq_lagrange_condition_optimal}
\frac{\partial L}{\partial \mathbf{w}}=0
\ \Leftrightarrow\  2\mathbf{\Sigma^{-}} \mathbf{w} - \lambda \mathbf{b}=0
\ \Leftrightarrow\  \mathbf{w}=\frac{1}{2} \lambda (\mathbf{\Sigma^{-}})^{-1} \mathbf{b}
\end{equation}
Note that the covariance matrix $\mathbf{\Sigma^{-}}$ is symmetric positive definite and so invertible. Adding the constraint, we obtain:
\begin{equation*}
\begin{split}
\mathbf{w}^{T}\mathbf{b}=1
\ &\Leftrightarrow\ \mathbf{b}^{T}\mathbf{w}=1
\ \Leftrightarrow\ \lambda \mathbf{b}^{T} (\mathbf{\Sigma^{-}})^{-1} \mathbf{b}=2\\
\ &\Leftrightarrow\ \lambda=\frac{2}{c_{rt}}\ \ \ \\
 \text{where} & \ \ c_{rt}=\mathbf{b}^{T}(\mathbf{\Sigma^{-}})^{-1}\mathbf{b}\ \ \text{is a constant.}
\end{split}
\end{equation*}
Finally, substituting in \refeq{eq_lagrange_condition_optimal}, we obtain the optimal weights $\mathbf{\widehat{w}^{-}}=\frac{1}{c_{rt}}(\mathbf{\Sigma^{-}})^{-1}\mathbf{b}$. Then, since $Var[\widehat{x}^{opt}_{RT}(t_{i})] = (\mathbf{\widehat{w}^{-}})^{T}\, \mathbf{\Sigma^{-}} \mathbf{\widehat{w}^{-}}$, we deduce the expression of the variance, which completes the proof.
\end{pfof}

~\\
\begin{pfof}{Lemme \ref{lemme_var_estim_PP}}
~\\
The estimators $\widehat{x}^{-}_{j}$ and $\widehat{x}^{+}_{j}$ are independent, so $Cov(\widehat{x}^{-}_{j}(t_{i}),\widehat{x}^{+}_{j}(t_{i}))=0$, and then:
\begin{equation*}
\begin{split}
& Var[\widehat{x}_{PP}(t_{i})]=  \sum_{j=1}^{N}\{(w^{-}_{j})^{2}\,Var[\widehat{x}^{-}_{j}(t_{i})]\\
& \qquad \qquad \qquad \quad \quad  +(w^{+}_{j})^{2}\,Var[\widehat{x}^{+}_{j}(t_{i})]\} \\
                         & +2\sum_{1\leq j<j'\leq N}\{w^{-}_{j}w^{-}_{j'}\,Cov(\widehat{x}^{-}_{j}(t_{i}),\widehat{x}^{-}_{j'}(t_{i}))\\
& \qquad \qquad \qquad +w^{+}_{j}w^{+}_{j'}\,Cov(\widehat{x}^{+}_{j}(t_{i}),\widehat{x}^{+}_{j'}(t_{i}))\}
\end{split}
\end{equation*}
with $\forall j=1,\ldots,N$ and $j<j'$, the expression of $Var[\widehat{x}^{-}_{j}(t_{i})]$ and $Cov(\widehat{x}^{-}_{j}(t_{i}),\widehat{x}^{-}_{j'}(t_{i}))$ have been proven in theorem \ref{th_var_estim_RT_asymp_nonrecurs}, $Var[\widehat{x}^{+}_{j}(t_{i})] \ =\ \sigma^{2}_{gps}+(j\lambda-d_{i})\,\sigma^{2}_{od}$ according to \refeq{eq_estim_plus}, and $Cov(\widehat{x}^{+}_{j}(t_{i}),\widehat{x}^{+}_{j'}(t_{i})) \ = \ Var[\sum_{k=i+1}^{g^{+}_{i}(j)}\varepsilon_{od,k}]
                                \ = \ (j\lambda-d_{i})\,\sigma^{2}_{od}$ by independence of $\varepsilon_{od,i}$.
\end{pfof}

~\\
\begin{pfof}{Theorem \ref{th_var_estim_PP_asymp}}
~\\
The estimator $\widetilde{x}_{PP}^{-}(t_{i})$  is equivalent to the estimator $\widehat{x}_{RT}^{\infty}(t_{i})$ defined in definition \ref{def_estim_RT_asymp_nonrecurs}. Thus, by symmetry, we deduce the expression of $\widetilde{x}_{PP}^{+}(t_{i})$ and then that the estimator $\widehat{x}_{PP}^{\infty}(t_{i})$ defined in \refeq{eq_estim_PP_asymp} is an asymptotically minimum variance unbiased estimator of the position of the vehicle at the sampling time $t_{i}$. Since $\sum_{j=1}^{N}\widetilde{w}_{j}^{-}=\sum_{j=1}^{N}\widetilde{w}_{j}^{+}=1$, we normalize $\widehat{x}_{PP}^{\infty}(t_{i})$ weighting by $(\widetilde{w}_{1},\widetilde{w}_{2})$ such that their sum is equal to one. The expression of weights $(\widetilde{w}_{1},\widetilde{w}_{2})$ are then deduce by the weighted least squares method. Finally, the expression of the variance of $\widetilde{x}_{PP}^{-}(t_{i})$ is derived from the equivalence between this estimator and $\widehat{x}_{RT}^{\infty}(t_{i})$. Thus, $Var[\widetilde{x}_{PP}^{-}(t_{i})]=Var[\widehat{x}_{RT}^{\infty}(t_{i})]$ whose expression is given in \refeq{eq_var_linear_estim_RT_asymp_nonrecurs}, and the expression of $Var[\widetilde{x}_{PP}^{+}(t_{i})]$ is deduced using lemma \ref{lemme_var_estim_PP}:
$Var[\widetilde{x}_{PP}^{+}(t_{i})]\ =\ (\widetilde{w}^{-})^{t}\, \Sigma^{+} \widetilde{w}^{-}
                                   \ =\ (\widetilde{w}^{-})^{t}\, \sigma_{gps}^{2}(I_{N}+r\, A_{N}(\lambda-d_{i}))\, \widetilde{w}^{-}$
which completes the proof.
\end{pfof}


\bibliographystyle{unsrt}
\bibliography{biblio_article_estimateur_position}

\end{document}